# Correlation of strontium anharmonicity with charge-lattice dynamics of the apical oxygens and their coupling to cuprate superconductivity.

Steven D. Conradson[a,b,*], Victor Velasco[c] Marcello B. Silva Neto[c,*], Chang-Qing Jin[d,e], Wen-Min Li[d,e], Li-Peng Cao[d,e]†, Andrea Gauzzi[f], Maarit Karppinen[g], Andrea Perali[h], Sandro Wimberger[i,j], Alan R. Bishop[k], Gianguido Baldinozzi[l], Matthew Latimer[m], Edmondo Gilioli[n]

[a]Department of Complex Matter, Josef Stefan Institute, 1000 Ljubljana, Slovenia

[b]Department of Chemistry, Washington State University, Pullman, WA 90164, U.S.A.

[c]Instituto de Física, Universidade Federal do Rio de Janeiro, Caixa Postal 68528, Rio de Janeiro, Brazil.

[d]Institute of Physics, Chinese Academy of Sciences, Beijing 100190, China.

[e]School of Physics, University of Chinese Academy of Sciences, Beijing 100190, China.

[f]IMPMC, Sorbonne Universites-UPMC, CNRS, IRD, and MNHN, Paris 75005, France.

[g]Department of Chemistry and Materials Science, Aalto University, Aalto FI-00076, Finland.

[h]School of Pharmacy, Physics Unit, Università di Camerino, 62032 Camerino, Italy.

[i]Dipartimento di Scienze Matematiche, Fisiche e Informatiche, Università di Parma, 43124 Parma, Italy.

[j]INFN, Sezione di Milano Bicocca, Gruppo Collegato di Parma, 43124 Parma, Italy

[k]Center for Nonlinear Studies, Los Alamos National Laboratory, Los Alamos, NM 87545, U.S.A.

[l]SPMS, CNRS CentraleSupelec Universite Paris-Saclay, Gif-sur-Yvette F-91192, France.

[m]Stanford Synchrotron Radiation Lightsource, SLAC National Accelerator Laboratory, Menlo Park, CA 94025, U.S.A.

[n]Institute of Materials for Electronics and Magnetism, CNR, Parma A-43124, Italy.

*to whom correspondence should be addressed: Steven D. Conradson, Marcello B. Silva Neto

**Email:** st3v3n.c0nradson@icloud.com, mbsn@if.ufrj.br

**Author Contributions:** Conceptualization-SDC, MBSN, CQJ, AG, MK, AP ARB; Methodology-SDC, MBSN, CQJ, AG, MK, GB; Investigation - SDC, MBSN, VV, WML, LPC, AP, SW, GB, ML; Visualization - SDC, MBSN, VV; Supervision - SDC, MBSN, CQJ, MK; Validation - AP, SW, ARB; Writing - original draft: SDC; Writing - review & editing: all authors.

**Abstract. Cu K edge** X-ray absorption spectra of overdoped superconducting $YSr_2Cu_{2.75}Mo_{0.25}O_{7.54}$ and $Sr_2CuO_{3.3}$ show a remarkably strong correlation of their superconductivity with the local dynamics of their Cu-Sr and Cu-apical-O pairs. This finding that the entire alkaline earth cation-apical O "dielectric" layer has an active role in the unusual electronic properties of cuprates has not been previously considered and has far reaching implications. We develop this idea of a possible role for the apical oxygen charge dynamics via a soft mode of the Sr by applying Kuramoto's synchronization technique to exact diagonalization calculations of two neighboring Cu-apical O pairs bridged by Sr and a planar O atom. These calculations show a first order phase transition to a synchronized state of the Internal Quantum Tunneling Polarons (IQTPs) in which a fraction of the hole originally confined to the apical O atoms of the cluster is transferred onto the planar O. This combination of experimental results and theory demonstrates that the Sr-O dielectric layer of cuprates most likely plays an important role in high temperature superconductivity via its collective charge dynamics that extends into the $CuO_2$ conducting planes .

## Introduction

Insofar as more than three decades of study of cuprates and related materials have not provided a consensus on the origins of high temperature superconductivity (HTSC) and related phenomena, there have been calls within the community for new ideas[1-4]. A source for these is found in recent reports on heavily overdoped superconducting compounds made with high pressure oxygen (HPO)[5-7]that describe exceptions to several of the common attributes of the more widely known cuprates doped by cation substitution or reaction with $O_2$ gas. These include: short Cu-apical oxygen (Oap) distances in compounds that retain high transition temperatures[8-10]; an oblate Cu geometry and inversion of the Cu $3d_{z2-r2}$ and $3d_{x2-y2}$ levels in $Ba_2CuO_{3.2}$[11]; superconductivity at 55-115 K at Cu valences far beyond the of ~2.27 limit of the "dome" of the conventional phase diagram[10]; the $CuO_{\sim1.5}$ plane with rotated magnetism in $Sr_2CuO_{3.3}$ (SCO)[12, 13]; and the first observation of a substantial transformation of the dynamic structure concomitant with the superconducting transition, found in SCO[14] and $CuBa_2Ca_3Cu_4O_{10+\delta}$[15, 16]. The extension of the superconductivity to excess charge of the $CuO_2$ plane Cu, *p*, to the maximum possible value of 1 demonstrates that the “dome” of superconductivity in the phase diagram only applies to the fraction of cuprates doped  by reaction with $O_2$ or cation substitution. The materials prepared with HPO therefore, obviate the presumed commonality of the widely accepted properties accruing to the dome and their basis for theories of HTSC.

Notably in the search for a universal characteristic, two-site distributions[17, 18] in the *dynamic* structures of Cu-apical O ($O_{ap}$) (or occasionally other Cu-O) pairs (Fig. 1a, b) coupled to HTSC have been observed in virtually all hole-doped cuprates via EXAFS measurements that probe the instantaneous structure factor[17-31], S(Q,t=0). This was corroborated by. neutron scattering measurements of the elastic structure factor[32, 33], S(Q,$\omega$=0) concomitantly with the original EXAFS results. Analogous behavior has also been found in non-cuprate HTSC compounds[34, 35]. Although we originally designated these as “tunneling polarons[20, 21, 36],” a more precise term differentiating them from Holstein-type and related small polarons is “Internal Quantum Tunneling Polarons” (IQTPs). IQTPs[21, 37, 38] occur when some of the excess charge of a Cu-centered small polaron is localized on one of its neighboring oxygen ions via, e.g., hybridization of the Zhang-Rice singlet state[39, 40]. This fractional decrease of its charge, $\delta$, modifies the Cu-O bond, causing a displacement of the oxygen from its crystallographic position to give a second, distinct Cu-O distance as occurs in the coordination chemistry of Cu-O where the chemical speciation determines the Cu geometry. It could follow that the longer copper-oxygen distance could be assigned to, formally, $O^{\delta-2}$ and the shorter to the original $O^{2-}$[21, 25]. The IQTP occurs when this charge and associated displacement shuttle back and forth between a pair of oxygen ions bonded to this same or adjacent copper ions via quantum (non-thermal) tunneling at a rate faster than the host polaron jumps to another Cu site, confining them to the O ions of this local cluster. Their essential characteristic is this fundamental internal charge-lattice dynamics constrained within a small number of neighboring atoms.

IQTPs become especially interesting from the perspective of the currently prevailing idea that the HTSC driver is primarily electron-electron interactions[41-45] augmented by the phonons found to be strongly coupled to it in ARPES[46-51] and other measurements[52-59] and calculations[60-62]. We have shown that at a critical value of an anharmonic coupling between the oxygen ions the pairwise exchange of the charge and displacement synchronizes, merging the wave functions to encompass the entire system[63, 64]. As a form of direct electron-phonon coupling in which intrinsic quantum oscillations of the atoms between their two sites constitute lattice-assisted dynamic charge transfer, the electron-electron interactions and entanglement that are a consequence of the in-phase oscillations of the Cu-O pairs of the IQTPs are a conceptual basis for the role of the lattice.,Whether the lattice is involved in the mechanism beyond simply providing a platform for the carriers has been one of the principal issues in HTSC[42] since shortly after its discovery.

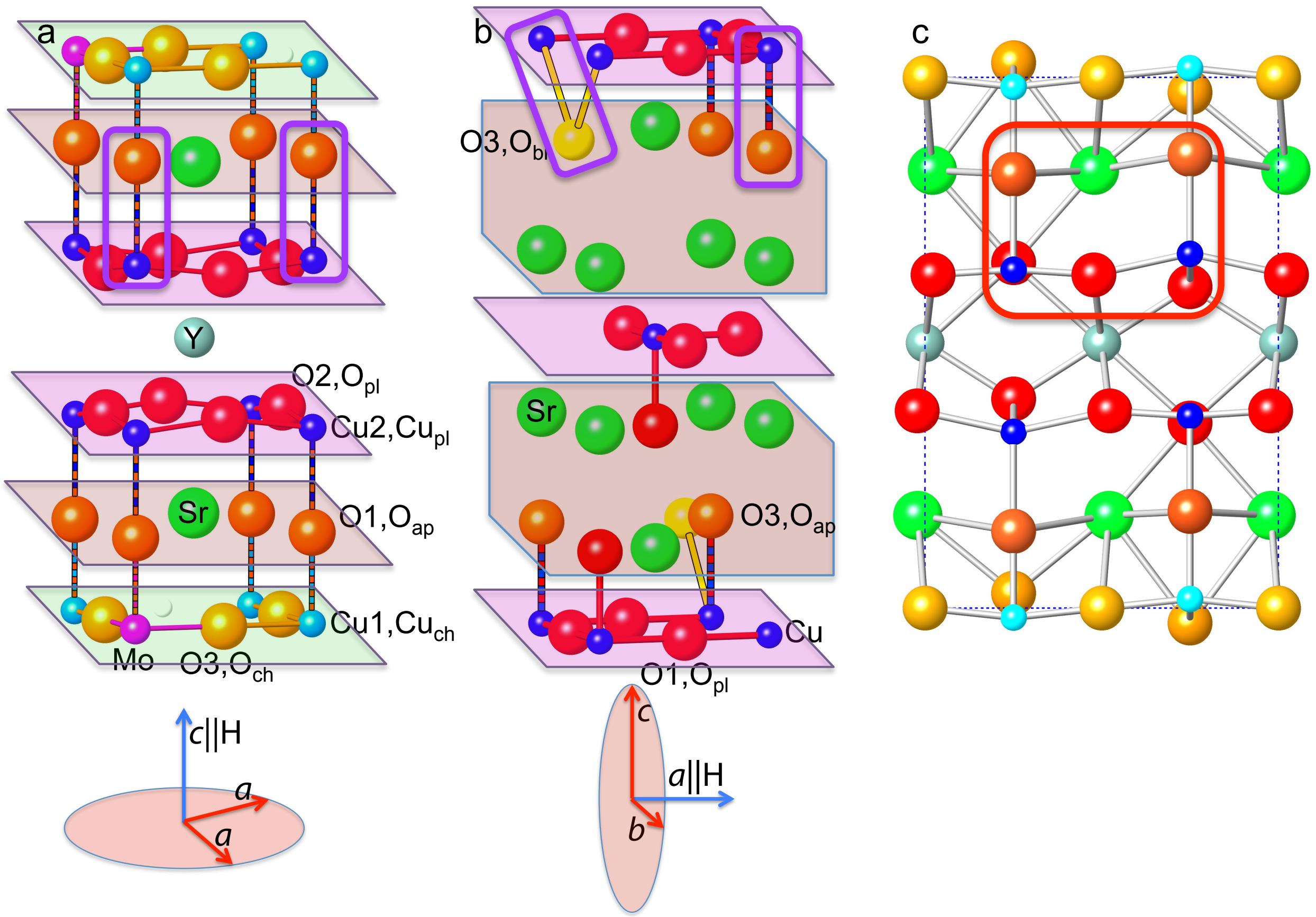

Figure 1. Structures of a) YSCO-Mo and b) SCO derived from crystallography and EXAFS, using the standard atom labels from the former. The directions of alignment in a magnetic field are shown at the bottom. The functional domains are: violet = conducting $CuO_2$ planes; green = charge reservoir layer for YSCO-Mo; beige = dielectric layer that for HPO SCO contains Sr and Oap but in conventional, divalent cation-doped $La_2CuO_4$-type compounds is composed of La and the Ca/Sr/Ba. The Mo environment becomes octahedral by incorporating the excess O into neighboring vacancies, introducing a fifth O around adjacent Cu ions. For SCO, half of the O sites between Cu ions along the *a* direction and many of the Oap sites are vacant, In its dynamic structure we have posulated that some Oap tilt towards adjacent Oap vacancies. Cu-O pairs that will form the IQTPs are outlined in purple. c) The symmetry-allowed atom displacements of the "X2-(a,0)" zone-boundary phonon for YSCO-Mo described below. The displacement magnitudes are arbitrary. The calculations also discussed in the next section were performed on the six atom cluster within the red rectangle.

IQTPs are identified in inelastic neutron pair distribution function (pdf) analysis[32, 65], while their coupling to,[17] and role in,[30] HTSC are elucidated by Extended X-ray Absorption Fine Structure (EXAFS). Here, we exploit the capabilities of EXAFS to probe the Cu-Sr pairs and extract changes in their distributions with high precision to advance our findings on the IQTPs in $YSr_2Cu_{2.75}Mo_{0.25}O_{7.54}$ (YSCO-Mo, $T_c$=84 K, Fig. 1) and SCO, $T_c$=95 K[12, 13]. These data are conjugate to vibrational spectroscopy that probes vibrational states via the energies of the collective displacements of atoms in a lattice. Alternatively, the EXAFS signal contains a snapshot of the real space distribution of specific pairs averaged over the probed volume. Anharmonic deviations from Gaussian-like distributions that soften and distort IR and Raman peaks[66] are manifested in the EXAFS as anomalously shaped increases in the normal exponential damping of its pairwise amplitudes that substantially diminish the magnitude of their Fourier transforms, $\chi(R)$[22].

An active role for the $O_{ap}$ atoms in HTSC has been demonstrated by numerous experiments and calculations. A possibility that has been mostly[67] neglected as unlikely because of the absence of any experimental data and the inert charge of the alkaline earth cations is analogous functionality for the entire Ba/Sr-$O_{ap}$ plane in the $YBa_2Cu_3O_{7\pm\delta}$-type (YBCO) and Bi/Hb/Tl-based compounds and the $La_{1-x}Ba/Sr_x$-$O_{ap=}$ layer that is two atoms thick in the $La_2CuO_4$-type (LCO) compounds. This is often referred to as the "dielectric" domain whose function has been presumed to be passively maintaining the separation of the excess holes injected into the $CuO_2$ "conducting layers" from the "charge reservoir" layer where they originate[68] (Fig. 1). Our recent reports on the EXAFS of YSCO-Mo and SCO focused on the Cu-O pairs and their IQTPs, but also showed that $\chi(R)$ from the Cu-Sr pair in YSCO-Mo along the *c* axis began to decrease at $T_c$. To further investigate this correlation we have now performed a much more thorough analysis of this component of the EXAFS. We find are that the Cu-Sr pairs are not only unusually soft, extending previous reports on their disorder[12, 69], but surprisingly they are also highly correlated with both the HTSC and the IQTPs. These results indicate that the entire dielectric domain, the cations as well as the $O_{ap}$ atoms, is highly active in HTSC via the local dynamical behavior of its constituent atoms. In addition, we have performed exact quantum diagonalization calculations on subunit of the structure containing these atoms that point to a possible mechanism for this process[63, 64]. These calculations elucidate the functionality of IQTPs that would underlie an important role for the dynamic properties of these entities in HTSC.

## Materials and Methods

The preparation and characterization of the samples and the transmission XAFS measurements at the Cu K edge at end station 2-2 of the Stanford Synchrotron Radiation Lightsource and analysis have been described previously in our reports on the local structure at a single temperatures[10, 13] and the near neighbor Cu-O pairs over a range of temperatures through the HTSC transition[14]. The calculation methods have also been described in detail[63, 64].

## Results

EXAFS measurements of the two HPO cuprates were performed to probe their dynamic structure and its coupling to the SC. Descriptions of the materials, sample preparation, and experiments have previously been reported[10, 13, 14]. These previous reports also described the behavior of the Cu-O pairs of the IQTPs in detail, but with only the brief description of the Cu-Sr in YSCO-Mo The novel aspect of this report is the detailed comparative analysis of the IQTP and Sr contributions to the EXAFS over a wide temperature range through the superconducting transitions in both compounds and its ramifications vis-à-vis our recent calcuations. This identifies the surprising strength of the correlation and corresponding coupling to the projection of the Cu-Sr dynamics along the *c*-axis.

This analysis is performed via direct visualization of the isolated contributions of the target Cu-O and Cu-Sr pairs under the assumption that the numbers of neighbor atoms remain constant, i.e., structural transformations are limited to displacements that alter distances but do not move atoms out of their original local environments. Changes in the distributions on the target pairs caused by either shifts of O atoms between different components of their multisite distributions or originating in ther local, pairwise dynamics are easily identified in the amplitudes of the Fourier transforms of their EXAFS, $\chi(R)$. Since multifrequency Fourier transforms of altered sine waves over limited ranges are not linear with respect to overlapping peaks it is necessary to separate the contributions of the different neighbor shells. This is especially important for smaller features that overlap larger ones. The first step in this isolation process is curve-fitting that obtains the forms of the individual waves that sum to the total EXAFS[10],. Curve-fitting also validates the structural model, finding the static neighbor shells that conform to the crystal structure and the dynamic ones that may deviate from it. The contribution of the target pair in $\chi(R)$ is then obtained by subtracting from the EXAFS these waves from those of the other shells of the structure. This process then expands the interpretation from a table of the metrical results from curve-fits to include a more qualitative visualization of correlations. Analogous to amplitude-ratioing analysis[70-73], it exploits the extraordinary sensitivity of EXAFS to changes in the pair distributions of closely related samples, e.g., temperature dependence and substitutions. Not only are the errors in the calculated phases and amplitude and backgrounds cancelled, but also the inability to determine exactly disordered structures and unknown pair distributions is rendered unimportant because the objective is to identify changes. Even if there are errors in the calculated waves that render the metrical data incomplete or with large uncertainties, the essential aspect is the reduction of the contributions of these errors to a level below that where it interferes with the spectral component of interest.

Another critical factor is the experimental resolution and the confidence that the signal from the target neighbor shell is not affected by and correlated with those of the neighbors that are being subtracted[72, 73]. This becomes especially important in isolating a smaller signal from a larger one with which it overlaps. The resolution of EXAFS is defined as $\Delta R=\pi/2k_{max}$ where $k_{max}$= the highest energy in the analysis[72]; the signature of two shells is the beat in the phase that differs in shape from simple damping from a broad anharmonic distribution. A not uncommon misconception is to relate the resolution to the width of the Fourier transform modulus peaks, in which case the degree of correlation in the signals between neirghboring shells would follow from the overlap of their peaks in $\chi(R)$. However, the few to several tenths of an Å width of the (R) peaks is a function of the finite and relatively short total range of the data used in their calculation; the waves for 1.9 Å Cu-O distances complete only six full cycles by k=15 $Å^{-1}$. That the FT widths are largely an artifact of their calculation is is also demonstrated by the use of window functions applied to the $\chi(k)$ spectra that eliminate ripple while broadening the features. All $\chi(R)$ data presented here were calculated with a sine window. For these spectra extending to k=15 $Å^{-1}$, 0.15 Å separations of the IQTP O shells give a beat at 11.5 $Å^{-1}$ that is completely within the spectra. Potential correlation between sets of shells from different elements, e.g., Sr and O, is further reduced by their different amplitudes and phase shifts.

This disconnect between the actual pair distribution and the shape of its feature in $\chi(R)$ is the interpretation of the behavior of the peaks in terms of changes in the distribution. In particular, this poses the question of the cause of the observable, which is amplitude reduction, which in the extreme case results in “spectroscopically” invisible neighbor shells. This originates in the limited range at both ends of the $\chi(k)$ spectrum, the inability to determine the amplitude at zero momentum transfer that defines the numbers of atoms, and the limited high energy range that results in the overlap of the contributions in $\chi(R)$ and the nonlinear summation of overlapping peaks even when their separation is much greater than the resolution.

*EXAFS of YSCO-Mo* The Fourier transform moduli of the original Cu $\chi(k)$ EXAFS, $\chi(R)$, of the total structure, the Cu2-Oap, and *c*-oriented projection of the Cu-Sr pairs of YSCO-Mo from 53 to 110 K are shown in Fig. 2. These were obtained as described above. The $k^3$-weighted spectrum was curve-fit with the Cu1-Oap, a single Cu2-$O_{ap}$, Cu2-Y, Cu2-Sr, and Cu1-Cu2 neighbors. The curve-fits did not find the Cu1-Sr and Cu2-Cu2 waves, indicating that these pairs are so disordered that they meet the “spectroscopically silent” criteria. This disorder of the Cu2-Cu2 pair is most likely thermal because of the length and presumed poor correlation of the Cu-O-Y-O-Cu link. The Cu1-Ba behavior would be a more extreme extension of our findings for the Cu2-Ba pair that we discuss below. All but the Cu2-$O_{ap}$ waves were then subtracted from the total spectra. The residuals were then Fourier transformed to give the results in Fig. 2a, with the flatness of the regions R>2 Å corroborating the negligible Cu1-Sr and Cu2-Cu2 contributions. These spectra have been presented before as overlays of the moduli. However, patterns and trends are more easily identified in these three-dimensional depictions, i.e., the absence of any anomalies in the major spectral features and very small change on the high R side of the O peak. This is especially true for the Cu2-$O_{ap}$ $\chi(R)$ (Fig. 2b). Above $T_c$ the peak at R~1.8 Å that is the contribution of the $O_{ap}$ with the shorter Cu-O distance is highest. At $T_c$ there is an abrupt drop in its amplitude, followed at lower temperatures by a shift of spectral weight to the R~2.0 peak that is assigned to the longer $O_{ap}$.This plot highlights the extremely narrow width of the change at $T_c$ and the subsequent loss of distinct features over a range below it. A change of the $O_{ap}$ contribution from a double or structured modulus peak to a more symmetric one at the transition is similar to the behavoir in other cuprates, although in those cases the cause was identified as a decrease in the separation between the two sites. YSCO-Mo differs from these other compounds in that, instead of recovering the original spectrum there is an obvious shift of spectral weight signifying a shift of some of the $O_{ap}$ from its position closer to Cu2 to the longer Cu2-$O_{ap}$ distance.Another peak at R=1.4 Å exhibits a distinct increase at $T_c$. Since this is too short for an actual Cu-O pair, we interpret it as SC-coupled anharmonicity in the Cu1-Oap distribution that does not affect its main peak and is not captured by the fit of this pair with a Gaussian distribution. Similarly, the Cu2-Sr χ(R) is superior to the previous amplitude ratios in showing the onset of the slow amplitude reduction at $T_c$ followed by the accelerating decrease with decreasing temperature (Fig 2c).

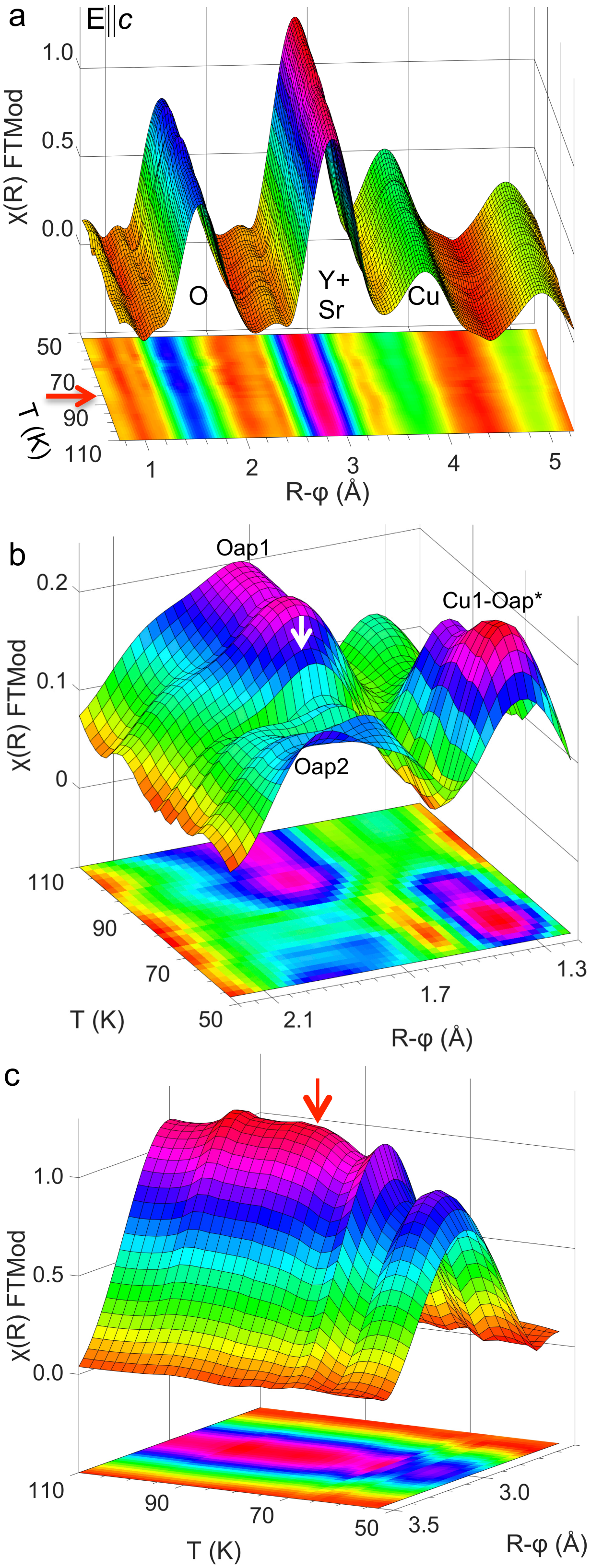

**Figure 2. E||*c* YSCO-Mo Cu EXAFS.** (a) Modulus of χ(R) of the total spectrum showing the neighbor atoms that are the origins of the spectral features. The arrow indicates the transition temperature=84 K. (b) Subtraction of the Cu1-$O_{ap}$, Y, Sr, and Cu-Cu waves from the spectra yields the χ(R) modulus of the two-site Cu2-Oap IQTP. The contributions of the two $O_{ap}$ sites to the broad or double peak are labeled 1 and 2. The Cu1-$O_{ap}$* peak is interpreted as the spectral contribution of an anharmonic component of the Cu1-$O_{ap}$ distribution after subtraction of the Cu1-$O_{ap}$ fit with a Gaussian distribution. (c) Cu2-Sr χ(R).modulus after isolation by same procedure.

The metrical data from from the curve-fits (Fig. 3) provide the additional details that make the essential connection. These were obtained by curve-fits of these residuals in which the sum of the numbers of oxygen atoms was fixed. The endpoint of k=14.7 $Å^{-1}$ gives a resolution of 0.11 Å. Insofar as the separation between the two O atoms is 0.17 Å, giving the beat in the composite wave in the EXAFS signifying the two Cu-O distances at k=9.3 Å, the Cu2-$O_{ap}$ χ(R) is not only asymmetric and and flattened across its top but shows two peaks when they are similar in amplitude. The Cu1-$O_{ap}$ distance that is the origin of the large O peak in Fig. 2a is 0.19 Å shorter than the shorter of the Cu2-$O_{ap}$ pairs, and the Cu1-Sr is more than 1 Å longer. Being substantially larger than the resolution, the separation is validated. Similarly, although the difference between the Cu2-Y and Cu2-Sr distances is just under the limit and their atomic numbers differ by only one, comparison of their waves from the curve-fits shows that they begin in phase at low energy, as expected, but are π out of phase by the higher energy limit. Insofar as the crystal structure dictates two distinct shells and not a single wide one, the correlation between their signals is therefore negligible and the isolation procedure is effective for the Sr as well.

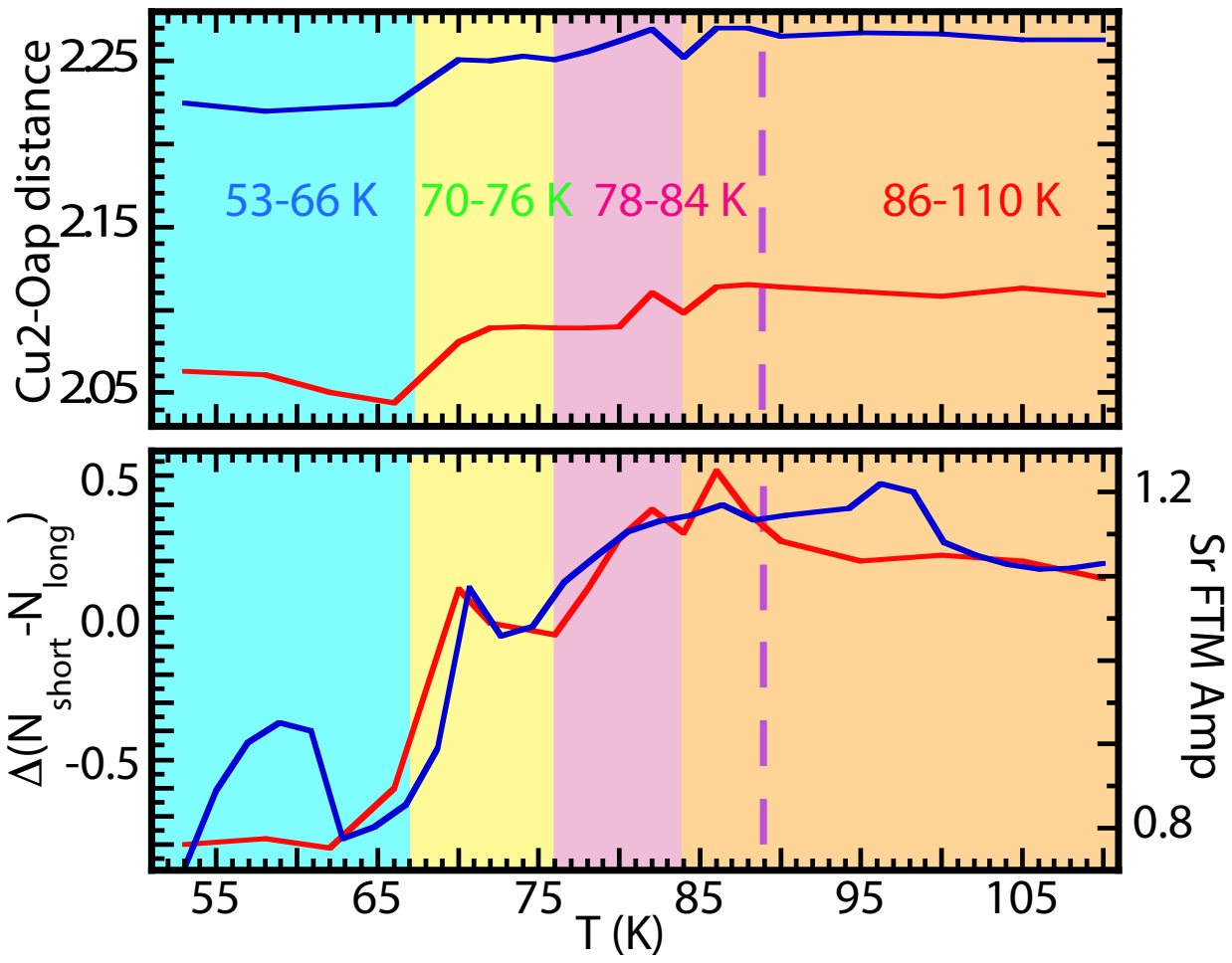

**Figure 3. Results for YSCO-Mo.** Top graph shows the two independent Cu-Oap distances calculated by curve-fits. They contract in parallel with decreasing temperature. The red trace in the lower graph shows the difference between the numbers of atoms at these shorter and longer distances. The blue, yellow, and orange zones delineate temperature ranges where the distances are constant within standard deviations less than the differences between them. Pink is a fluctuation region that begins at $T_c$ for the distances and somewhat higher (the dashed line) for the numbers of atoms. The blue trace is the magnitude of the *c*-oriented Cu2-Sr χ(R).

Our prior analysis of the Cu-$O_{ap}$ pairs found that, based on the two independent Cu-O distances and numbers of atoms, the IQTPs exhibit identical patterns of plateaus and discontinuities at the same temperatures. These delineate the normal phase, a fluctuation region at the transition, and two separate regions in the superconducting phase. Reproducing them here with the addition of the χ(R) to the difference in the numbers of oxygen atoms accentuates the novel finding: the peak of the Sr χ(R) displays a remarkable correspondence with the shifts of the two O atoms between their two sites. This even includes the feature at 70 K that would be assigned to a noise spike if it only occurred in one of these results. We emphasize that nothing in the above data analysis would bias this correlation between these three parameters and the superconducting transition. The *c*-oriented χ(R) peak for the Cu2-Sr (Fig. 2c) is relatively flat at higher temperature, and then begins a slow curve downward at $T_c$ that subsequently drops more steeply, albeit with some additional features. This behavior is indicative of increasing disorder with decreasing temperature, the opposite to the conventional thermal broadening of the distribution. The correlation is with the difference in the populations of the two Cu2-$O_{ap}$ sites caused by changes in the relative energies of the two minima of the potential (Fig. 3). The high level of correlation demonstrates the direct coupling of the anharmonicity of the Cu2-Sr distribution to the renormalization of the Cu2-$O_{ap}$ potential and both of these to the SC, including the division into the four regions.

*XAFS of SCO.* Although its composition and parent structure are simple, the EXAFS and local structure of our second material, SCO, are much more complicated than those of YSCO-Mo. A notable aspect is the large changes in the amplitudes with temperature. These are often opposite to the normal reduction with increasing temperature, and several occur over narrow ranges both in proximity to and away from the HTSC transition. Also remarkable are the significant features between the principal O and Sr peaks that do not correspond to sites in the crystallographic structure (Fig. 1) that essentially disappear at $T_c$. Our earlier study of SCO found that it is unique among cuprates in exhibiting oxygen vacancies not only in its apical positions but also in its $CuO_2$ planes. That the planar vacancies occupy approximately half the sites on only one axis provides an explanation for why its alignment in a magnetic field is along that direction instead of its *c* axis. Because of this rotated magnetism, the Cu-O IQTP in SCO is observed in its crystallographic *bc* plane[13]. Another unique feature is the magnitude of the change in its dynamic structure through its superconducting transition (Fig. 4a).

We originally postulated that the behavior of the spectra across the transition might include a Sr atom with a Cu-Sr distance less than 3 Å[14]. However, visualizing the affected spectral region by the isolation method shows for the first time that the large amplitude features at R=2.2-2.8 Å are present above as well as below the transition. The broad, flat top of this feature from 85-95 K (Fig. 4b) causes it to be assigned to the two-site Cu-O distribution characteristic of IQTPs. The remarkable result is its disappearance at $T_c$ signifying a radical change in the dynamic structure coupled to the superconducting transition, a finding not only outside of BCS SC but also for HTSC. The absence of a correlation between the amplitudes of the two contributions to the peak is also unique. The current best model for this behavior is that the IQTP oxygen atoms are most likely apical, next to an apical vacancy, displaced along the *b* axis to asymmetrically bridge two Cu atoms at ~2.66 and 2.91 Å. Both these distances and the 0.25 Å separation are much longer than any other cuprates. The radical difference in the Cu geometries relative to other cuprates, including $O_{pl}$-Cu-$O_{ap}$ angles lower than 90°, could be a consequence of the much higher Cu valence in this overdoped compound. The oxygen atom below them in the *b*-oriented Cu-O chain shifts away from this $O_{ap}$ to slightly expand its Cu-O bonds[14]. The IQTP could then involve tunneling of the oxygen atoms into equivalent sites across the *ac* plane that passes through the midpoint of the two coppers. χ(R) at 121 K in the normal phase has most of its amplitude in the peak for the longer Cu-O distance, with the shorter oxygen giving just a shoulder. The almost complete loss of amplitude at $T_c$ could result from either the separation between the two sites decreasing to maximize the interference of their EXAFS waves and/or their distributions broadening in the fluctuation region of the transition. Below $T_c$ the two contributions from the two sites are equal. They begin to fall off at ~80 K, with the higher R peak again being larger. A possible explanation for the diminished overall amplitude of the IQTP at lower temperature is that in this postulated configuration for SCO the oxygen could shift to a normal $O_{ap}$ position directly above the Cu along the *c* direction.

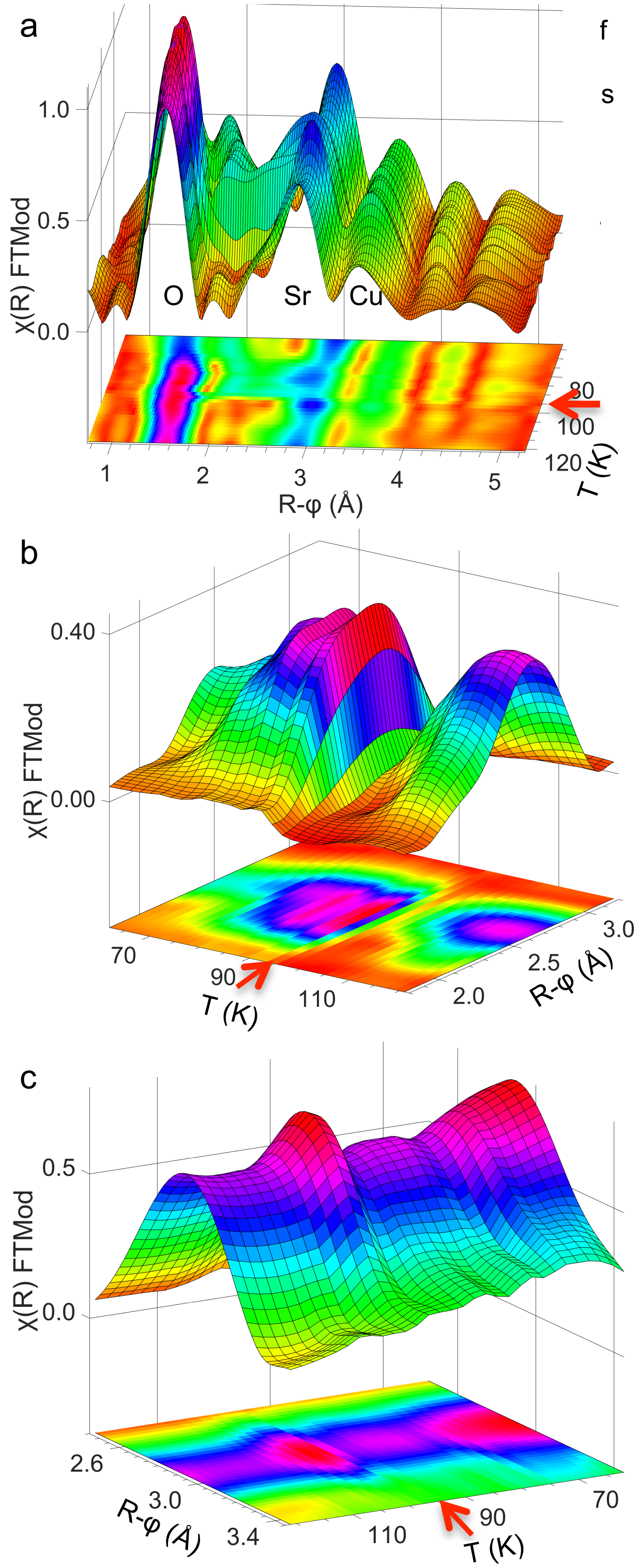

**Figure 4. E||*bc* SCO Cu EXAFS.** (a) Modulus of χ(R) of the total spectrum labeling the neighbor atoms that are the origins of the spectral features assigned to crystallographic sites. The arrow indicates the transition temperature=95 K. (b) Subtraction of the Cu-O, Sr, and Cu-Cu waves from the spectra yields the χ(R) modulus of the two-site Cu-Oap IQTP as described in the text. The contributions of the two Oap sites are evident in the broad peak below $T_c$, although they do not track together except for the loss of the total feature just above $T_c$. (c) Cu-Sr χ(R) modulus after isolation by same procedure.

The Cu-Sr contributions were fit with two Sr neighbors at ~3.23 and 3.36 Å. This 0.13-0.14 Å separation is only marginally larger than the resolution for the k=14.2 $Å^{-1}$ upper limit of the curve-fit. In combination with the disorder that also drives the amplitude down at high k, the increase in the amplitude after the beat will be minimal and the Cu-Sr waves in χ(k) will be indistinguishable from a single, highly damped one. What is clearly observed now that was obscured in our prior report is that χ(R) shows a single, round peak with minimal structure. The Cu-Sr χ(R) amplitude rises as the temperature is lowered from ~78 K, also developing a more distinct shoulder on the high R side, where the second Sr at the longer distance needed for a complete curve-fit makes its contribution. This relatively normal trend is broken by the peak beginning near 106 K that falls immediately at $T_c$, reflecting a narrowing and hardening of the Cu-Sr distribution in the fluctuation region that is eliminated on becoming superconducting. All of these changes can be presumed to originate in the shape and resultant ordering within the single peaked Cu-Sr distribution.

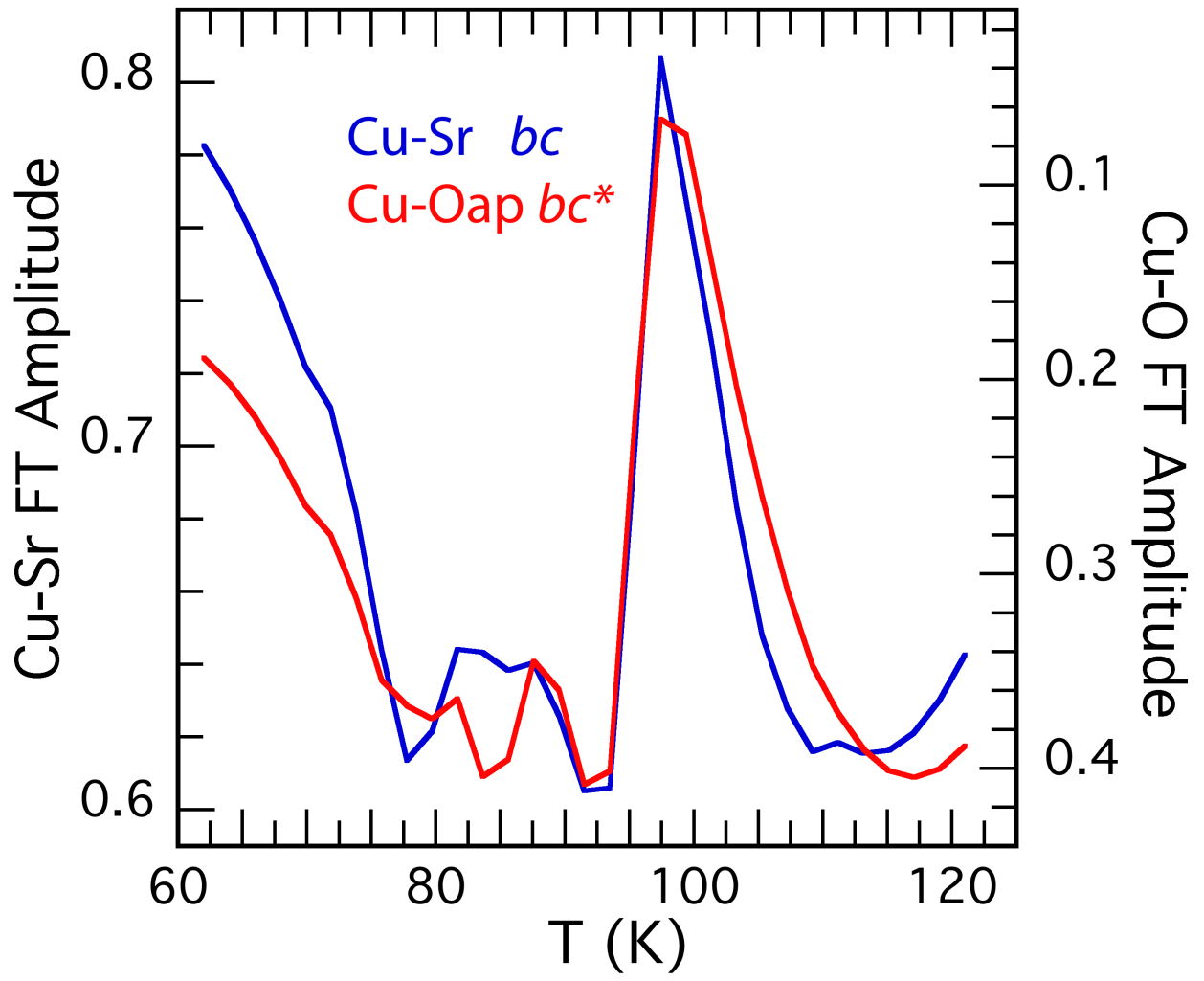

**Figure 5. E||*bc* SCO χ(R).** χ(R) modulus amplitudes of of the isolated Sr and R=2.6 Å IQTP (= *) contributions. The inverse behavior of the two features is apparent.

Because of the difficulty of curve-fitting to extract metrical parameters when the features have vanished from the spectra, the correlation with the Sr is derived from comparing the isolated χ(R) peaks from the IQTP O and Sr atoms at, respectively, R=2.6 and 2.95 Å. The novel result is the high degree of correlation, or in this case anti-correlation, between the Cu-Sr and the IQTP χ(R) (Fig. 5), most notably the dramatic response to the onset of the SC. As with the YSCO-Mo, the O atoms of the IQTP and the Sr are strongly coupled both to each other and the SC.

*Modeling with IQTP-coupled anharmonic Sr.* Our demonstration of an anomalous Cu-Sr distribution that is strongly correlated with both the dynamical structure of the Cu-$O_{ap}$ and HTSC poses the questions of the HTSC-lattice coupling and the interaction of the Sr with the IQTPs. We have explored this by applying exact quantum diagonalization[21, 37] to a six-atom cluster (Fig. 1c) that is an extension of the original, three atom O-Cu-O moiety previously used to elucidate the experimental signatures of IQTPs on the elastic and instantaneous structure factors. Briefly summarizing the results of the calculations emphasizes the importance of the strong anharmonic coupling of the Sr to HTSC found in EXAFS and its possible or even likely role in the HTSC mechanism. The minimal subunit encompassing the relevant constituents consists of a pair of neighboring Cu-Oap IQTPs bridged between their Cu ions by an $O_{pl}$ and between their $O_{ap}$ by the Sr of the dielectric layer (Fig. 1)[63, 64]. The novel aspects enabled by this structure are the addition of the $CuO_2$ plane through the inclusion of $O_{pl}$ and the anharmonic structural dynamics of the Sr atom that is incorporated via the nontrivial, $O_{ap}$-Sr-$O_{ap}$ triatomic molecular structure indicated by these EXAFS results. A single extra hole initially localized on one of the two $O_{ap}$ atoms causes its displacement to give the two-site distribution. The starting point for the calculation was to consider this cluster as a classic Holstein-type polaron in which a significant fraction of the excess charge is found on the $O_{ap}$. The standard Hubbard-Holstein Hamiltonian used to describe the interplay between electronic and lattice degrees of freedom was augmented by an anharmonic phonon-phonon coupling of strength K describing the interaction between the triatomic O-Sr-O molecular vibration and the independent apical phonons located on the $O_{ap}$ sites. Reiterating from the introduction, the internal dynamics within this cluster defines the IQTP as the oscillation of the excess charge and displacement on an $O_{ap}$ site via tunneling at a frequency higher than thermally activated hopping of the polaron to a neighboring site that is accomplished by exchanging these characteristics with an adjacent, degenerate oxygen.

In addition to the numerically exact diagonalization calculation the charge-lattice dynamics was evaluated via the Kuramoto approach for synchronization of coupled oscillators by equating the anharmonic coupling K of the the O-Sr-O vibration to the coupling between phase oscillators in the Kuramoto model[74,

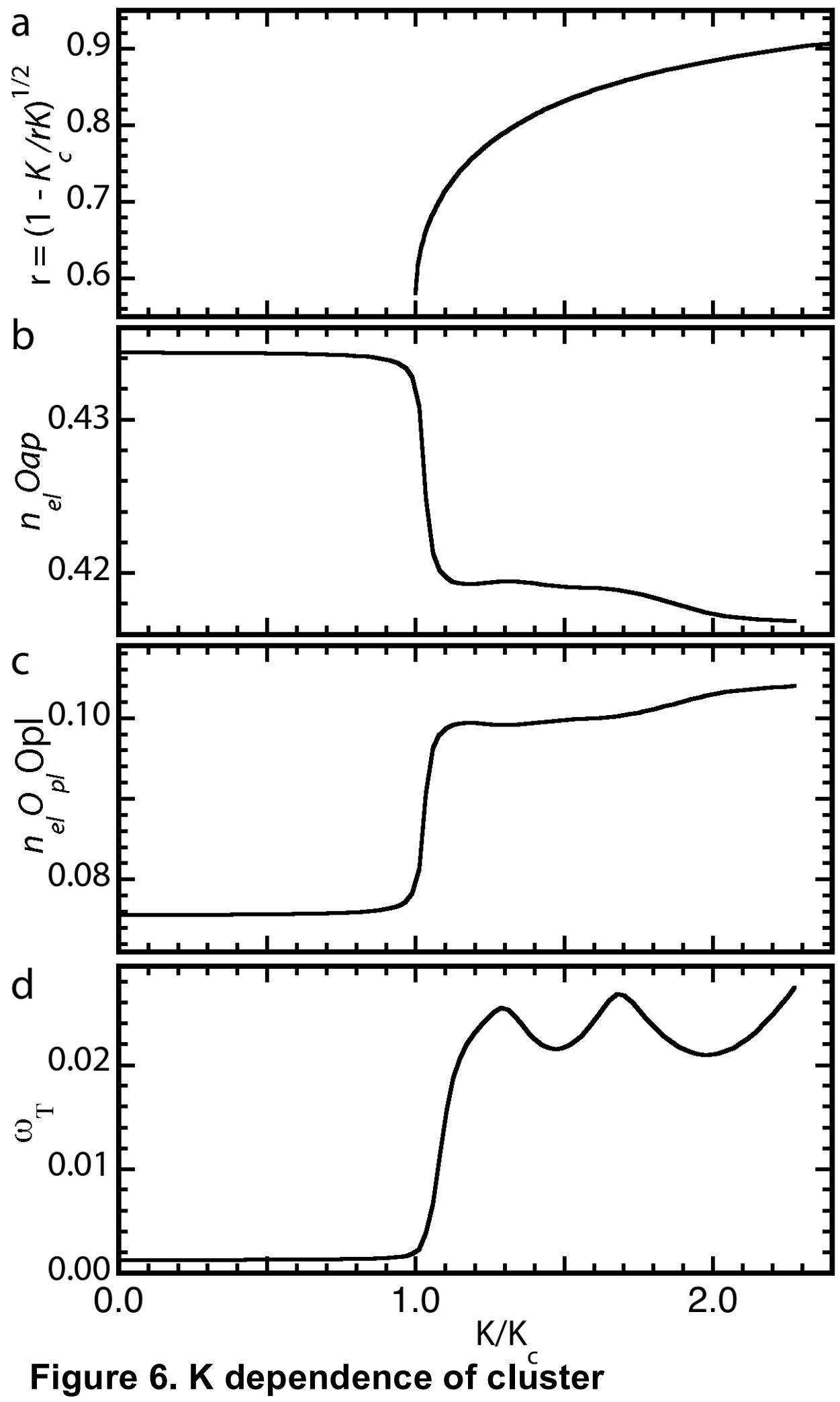

**Figure 6. K dependence of cluster characteristics.** a) O-Sr-O phonon order parameter. b) excess hole density on $O_{ap}$. c) excess hole density on $O_{pl}$. d) tunneling frequency.

[75]. In addition to its application here incorporating the EXAFS reults, this approach has also been applied to the emergence of superconductivity as a synchronization problem[76]. The calculations shown here were performed by varying the anharmonic interaction strength, K, while fixing the electron-phonon coupling, λ, in the non-adiabtic regime that defines the cluster. The diagonalization procedure showed that at a critical value of K, $K_c$, the system undergoes a first order transition defined by, discontinuities in e.g., the phonon order parameter, the charges on the oxygen ions, the tunneling frequency, etc. (Fig.6). The Kuramoto analysis showed that this transition can be understood as a dynamic synchronization; when $K>K_c$ correlated, antiphase tunneling of the two Cu-$O_{ap}$ pairs is initiated between their opposite configurations of displacement and charge. Analogous calculations have shown that the formation of the synchronized phase does not occur with a harmonic bridging mode, the anharmonicity found by EXAFS is mandatory[77, 78].

For $K<K_c$ that is the unsynchronized phase the excess charge is localized on either the left or right $O_{ap}$, forming two separated potential wells (Fig. 7a), between which no tunneling occurs. When the system transforms to the synchronized, phase locked state the resulting delocalization of the charge transfers a fraction of the excess charge originally on the two $O_{ap}$ atoms to $O_{pl}$. The synchronization extends the wavefunction to form this third local minimum on $O_{pl}$ (Fig. 7a) and collapses the energy levels so that the excitation energies are greatly reduced. This delocalization and dispersion throughout the cluster of the hole that started on a single $O_{ap}$ is a clear signature that the synchronization of the IQTPs through the anharmonic $O_{ap}$-Sr-$O_{ap}$ phonon modifies the electronic properties of the entire system. The resulting triple-well functions as an anharmonic-structural-adiabatic-passage promoting charge delocalization, that once formed may further enhance the oscillator synchronization in a positive feedback loop. Using the example of the polaronic tunneling frequency, $\hbar\omega_T = E_1 - E_0$ (Fig. 6d), that is the difference in energy between the ground and first excited states, it is seen that at K values below the critical value $K_c$ $\hbar\omega_T = 0$, no tunneling occurs, and the polarons are trapped on their original left or right sites. Above $K_c$ the abrupt increase in $\hbar\omega_T$ is the signature of the transition to the synchronized phase. Because it is essentially a spectroscopic parameter, an interesting aspect of $\omega_T$ is the structure in the synchronized phase above $K_c$ (Fig. 6d) that is caused avoided crossings in the ground to excited state transition. Such structure does not occur in the other parameters.

## Discussion

These results provide a critical extension to our prior EXAFS study that demonstrated that the two-site Cu-O distributions in the dynamic structure that constitute the IQTPs are strongly coupled to the HTSC[17, 19, 23 24 30] in overdoped, superconducting SCO and YSCO-Mo[14]. One immediate issue raised by these two compounds is the conservation or even enhancement of their HTSC after the insertion of the excess oxygen atoms by the HPO process that raises the charge on the Cu to values far outside the "dome" of the conventional phase diagram[79]. Whereas this additional oxygen raises $T_c$ by~1/3 relative to the 63 K of parent $YSr_2Cu_3O_{7+\delta}$, a vlaue still somewhat under the maximum of the YBCO family, the radically different locations of the adventitious oxygen and disruption of the $CuO_2$ planes occur comcomitantly with a $T_c$ of SCO that is more than 2-1/2 times that of its parent $La_2CuO_{4+\delta}$. The loss of HTSC with overdoping by reaction with $O_2$ or cation substitution that closes the dome on its high *p* side[80, 81] therefore cannot result solely from saturation of the $CuO_2$ planes with holes[79] but must be at least partly a lattice effect specific to those structures. The quantum phase diagram without such lattice effects is the one for HPO cuprates[7, 10] where the HTSC extends to the highest attainable doping levels with retention of the transition temperature and superconducting volume. We point out that HTSC HPO compounds of both the YBCO and LSCO classes were first made shortly after the initial discovery of HTSC[15, 82-84], but despite ongoing work[6, 7, 9, 85, 86] have been neglected by the community.

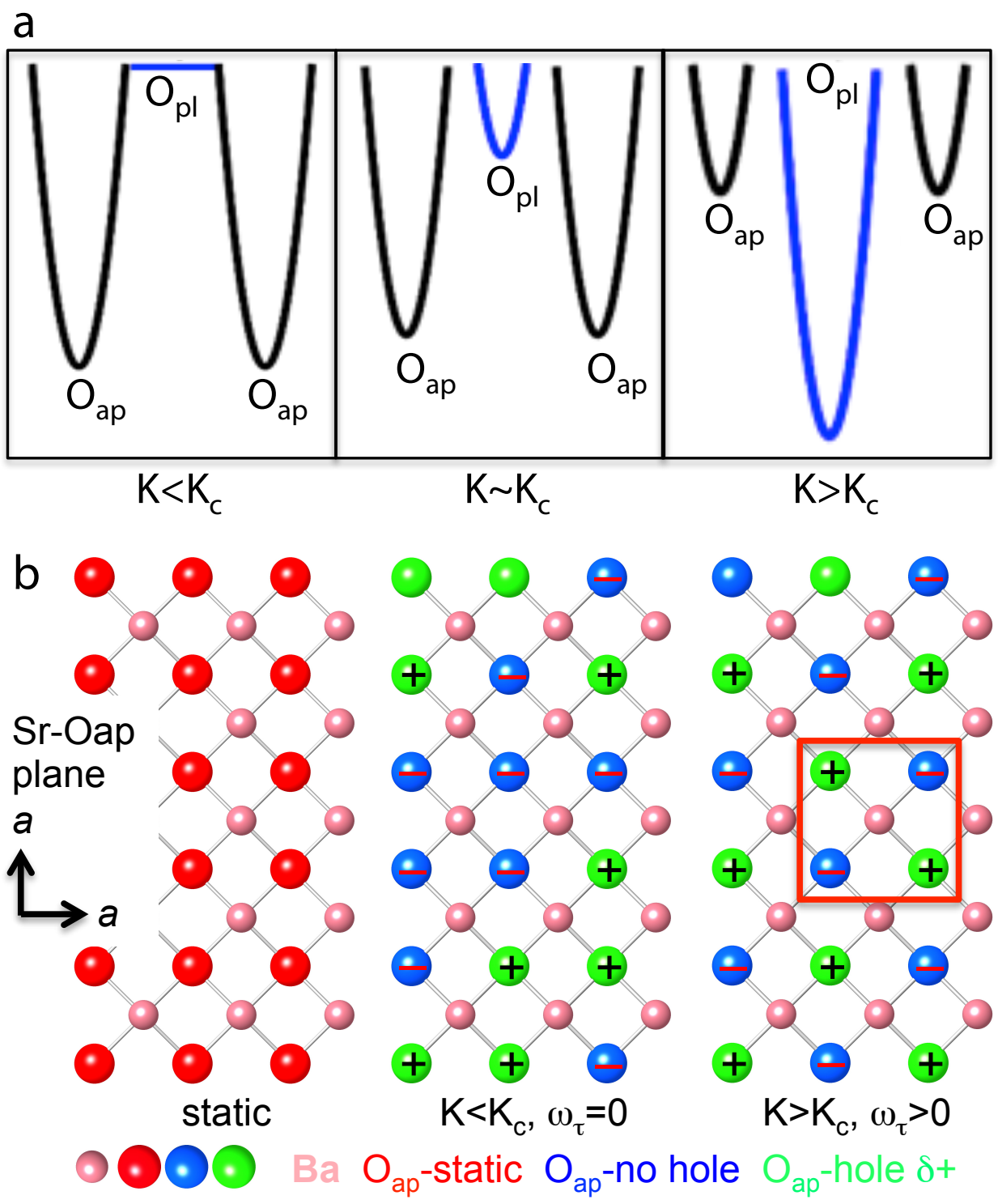

**Figure 7. Charge distributions.** a) Density of excess hole on the two $O_{ap}$ and $O_{pl}$ in the three regimes of the phase diagram for different K values. Prior to synchronization the hole resides on either one of the other of the two $O_{ap}$ atoms. As K increases, starting at the to initiation of the IQTP synchronization at K~$K_c$, increasing amounts of hole density are transferred to $O_{pl}$. b) Schematic of dielectric, Sr-$O_{ap}$ layer looking along *c* direction. In the static structure all of the $O_{ap}$ are in the same crystallographic postion. In the instantaneous or dynamic structure for K<$K_c$, in this snapshot half of the oxygen atoms are at the site with the longer Cu-O distance (green, "+") and the other half exhibit the shorter Cu-O distance (blue, "-"), which are randomly distributed. For K>$K_c$ in the synchronized phase the ratio is the same but now a snapshot of the $O_{ap}$ atoms shows alternating up and down to double the unit cell size of the dynamic structure. This structure exhibits *d* symmetry about the Sr ion at the center of the $CuO_2$ squares.

In addition to the challenges to existing theories posed by the HPO phase diagram, these differences between YSCO-Mo and SCO that reinforce their common factor being strong coupling of the anharmonic, dynamic structure of Cu-Sr and IQTPs to each other and the HTSC takes on added significance. These new EXAFS measurements greatly extend the range of the structure and behavior of IQTPs beyond the previously observed partial collapse of the double well potential near the HTSC transition. Not only are the two minima of the double well potential non-degenerate, but also the difference in their energies varies across the entire temperature range with a large shift at the superconducting transition even while their separation is unchanged. The essential factor in the HTSC-IQTP coupling is therefore not limited to the equal splitting in the dynamic structure of the $O_{ap}$ sites into a two-site distribution along the *c* axis with tunneling frequencies of the order of 100 $cm^{-1}$. The coupling is enlarged to include alternative double well potentials and their associated charge dynamics where the two energies, locations in the lattice, and response to the HTSC transition can vary.

The novel extension of the IQTP is the varied anharmonicity and coupling of the $O_{ap}$ of the IQTPs to the $Sr^{2+}$ ion with which it constitutes the dielectric layer. Far from being passive[68], our calculations demonstrate that this coupling and consequent synchronized, antiphase motions of the $O_{ap}$ atoms and their associated charges have substantial effects on the dynamics and charge distributions in the other domains of the materials, notably the $CuO_2$ planes where the charge on the oxygen is highly correlated with $T_c$[87]. The extension of the localized wave function on $O_{ap}$ to multiple IQTPs and sites in the $CuO_2$ plane promotes electron-electron interactions that go far beyond the acknowledged special role of the $O_{ap}$ atom in controlling the carrier density[87-89] or, in the extreme case, promoting pairing directly via lateral vibrations[67] (that are not observed in EXAFS). The synchronization proffered by the anharmonically coupled two-site distributions of the IQTPs therefore also entangles at least some of the relevant degrees of freedom between the conduction layers and the dielectric layers and charge reservoirs. The addition of a second IQTP pair, encompassing all four neighboring IQTPs within the unit cell (Fig. 1), requires a second nearest neighbor coupling constant, $K_2$, that describes the anharmonicity diagonally across the unit cell. Even so, in the trivial extension in the square lattice to larger clusters, or even to the thermodynamic limit of a large fraction of the crystal, the Kuramoto analysis persists in showing anti-phase synchronization between nearest- (anharmonicity $K_1$) and next-to-nearest- (anharmonicity $K_2$) IQTP neighbors. This indicates that the entanglement is not confined but instead encompasses the entire synchronized domain that is capable of occupying the square lattice of the entire crystal. Furthermore, mapping this phase dynamics onto a $K_1 - K_2$ model for dipolar oscillators gives a ground state configuration corresponding to an angular momentum $\ell = 2$ (Fig. 7b), i.e., *d* symmetry.

This mapping of the *d*-wave symmetry of the HTSC to that of certain phonons, e.g., the $B_{1g}$ phonon that is the antiphase buckling of the Cu-O-Cu moieties of the $CuO_2$ plane[60], was shown shortly after the discovery of HTSC[33, 90], although there is disagreement on the strength of its coupling to the HTSC[91]. Anomalies in phonon dispersions indicative of strong electron-phonon coupling that are correlated with the both $T_c$ and the doping, e.g., the "giant softening" of the bond-stretching phonon frequencies off of the zone center caused by strong renormalization along [010] and [100] and avoided crossings are a common characteristic of cuprates[53-56, 92]. Although these are still insufficient to HTSC, insofar as the doping-induced charge inhomogeneities are localized[93, 94] the widths of the phonon peaks reflect their resulting wide range of energies[95]. The comparison the *p*-dependence of these ~70 meV phonon dispersion anomalies with the "kink" in angle-resolved photoemission denotes different origins for these two spectral signatures. Whereas this feature in ARPES would indicate electron-anomalous phonon coupling, its presence in neutron scattering would result from collective charge excitations[56, 96] that would promote a novelthe HTSC mechanism.This would be an apt description of an important role for a synchronized IQTP phase. Although there are no zone center phonons with the antiphase displacements of the synchronized phase of the IQTPs, analogous to the neutron scattering the correlated Cu-$O_{ap}$ motions could originate from a phonon not belonging to the Brillouin zone centre, Utilizing the $YBa_2Cu_3O_7$ (YBCO)-type structure (Fig. 1c) the calculated atomic displacements are described by a symmetry-adapted coordinate belonging to the $X_2^-$ irreducible representation at the X point of the tetragonal Brillouin zone, where only one branch of the star (0 1/2 0) is involved. This off zone-center phonon exhibits the same *d*-symmetry with respect to the centers of the $CuO_2$ squares as found in the extension of our calculation results to neighboring IQTP pairs.

Reiterating, in a doped system with the appropriate combination of U, t, J, and relative energies of the M and O states to give the requisite mix of ionic and covalent character in the Cu-O bond, and that also has two species possessing M-O bonds separated by a barrier of requisite height and width, the IQTPs are a natural outcome of coupling this phonon to the coordination chemistry of the transition metal ions and the preferred geometries associated with their valence. The partial localization of the charge causes the metal sites to adopt the these geometries and bond lengths within the lattice. This will also result in the known sensitivity to the interatomic distances and angles[80, 81]. If the thermal motion of the Cu-O pair traverses this range then instead of the potential being featureless it can evolve into the two-site distribution by developing local minima at those locations corresponding to the preferred charge distributions and their bonding modes. The EXAFS results therefore point to at least this candidate for the formation of entangled, synchronized Cooper pairs and condensation in a *d*-wave channel. Additional corroboration of this conjecture is provided by recent momentum-resolved electron energy loss spectroscopy results that, with other measurements on the strange metal phase, attribute its scale invariance[58, 97] to a"highly entangled man-body state" containing "new collective nonlocal entities as the charge carriers" that would originate in the combined electron-electron interactions and electron phonon coupling that we find in synchronized IQTPs.

**Acknowledgments.**
Funding for this work was provided by: Slovenian Research Agency core funding P1-0040 (SDC); National Science Foundation grant 1928874 (SDC); Department of Energy, Office of Basic Energy Sciences contract DEAC02-76SF00515 (SDC, ML); Ministry of Science and Technology of China (CQJ, WML, LPC); Natural Science Foundation of China (CQJ, WML, LPC)
Department of Energy, Office of Basic Energy Sciences DEAC02-76SF00515 (SDC, ML).